\documentclass[%
  aps, a4paper, twocolumn, pra, superscriptaddress, amsmath, amssymb, floatfix
]{revtex4-1}
\usepackage{graphicx}
\usepackage{epstopdf}
\usepackage[table]{xcolor}
\usepackage{braket}
\usepackage[colorlinks,bookmarksopen=false]{hyperref}
\usepackage[caption=false]{subfig}
\usepackage{bbm}
\usepackage{lipsum}
\usepackage{siunitx}

\DeclareMathOperator{\Tr}{Tr}

\begin{document}

\title{%
 IBM Q Experience as a versatile experimental testbed \\ for simulating open quantum systems
}

\author{G. Garc\'{\i}a-P\'erez}

\affiliation{QTF Centre of Excellence, Turku Centre for Quantum Physics, Department of Physics and Astronomy, University
of Turku, FI-20014 Turun Yliopisto, Finland}

\affiliation{Complex Systems Research Group, Department of Mathematics and Statistics,
University of Turku, FI-20014 Turun Yliopisto, Finland}

\author{M. A. C. Rossi}

\affiliation{QTF Centre of Excellence, Turku Centre for Quantum Physics, Department of Physics and Astronomy, University
of Turku, FI-20014 Turun Yliopisto, Finland}

\author{S. Maniscalco}
\affiliation{QTF Centre of Excellence, Turku Centre for Quantum Physics, Department of Physics and Astronomy, University
of Turku, FI-20014 Turun Yliopisto, Finland}
\affiliation{QTF Centre of Excellence, Center for Quantum Engineering, Department of Applied Physics,
Aalto University School of Science, FIN-00076 Aalto, Finland}

\maketitle

{\bf The advent of Noisy Intermediate-Scale Quantum (NISQ) technology is changing rapidly the landscape and modality of research in quantum physics. NISQ  devices, such as the IBM Q Experience, have very recently proven their capability as experimental platforms accessible to everyone around the globe. Until now, IBM Q Experience processors have mostly been used for quantum computation and simulation of closed systems. Here we show that these devices are also able to implement a great variety of paradigmatic open quantum systems models, hence providing a robust and flexible testbed for open quantum systems theory. During the last decade an increasing number of experiments have successfully tackled the task of simulating open quantum systems in different platforms, from linear optics to trapped ions, from Nuclear Magnetic Resonance (NMR) to Cavity Quantum Electrodynamics. Generally, each individual experiment demonstrates a specific open quantum system model, or at most a specific class. Our main result is to prove the great versatility of the IBM Q Experience processors. Indeed, we experimentally implement one and two-qubit open quantum systems, both unital and non-unital dynamics, Markovian and non-Markovian evolutions. Moreover, we realise proof-of-principle reservoir engineering for entangled state generation, demonstrate collisional models, and verify revivals of quantum channel capacity and extractable work, caused by memory effects. All these results are obtained using IBM Q Experience processors publicly available and remotely accessible online.}

\section*{Introduction}
The theory of open quantum systems studies the dynamics of quantum systems interacting with their surroundings~\cite{BreuerPetruccione,Weiss,RivasHuelgaBook}. In its most general formulation it allows us to describe the out-of-equilibrium properties of quantum systems, it provides a theoretical framework to assess the quantum measurement problem, and it gives us the tools to investigate, understand and counter the deleterious effects of noise on quantum technologies. For this reason its range of applicability is extremely wide, from solid state physics to quantum field theory, from quantum chemistry and biology to quantum thermodynamics, from foundations of quantum theory to quantum technologies.

Generally, the dynamics of open quantum systems are described in terms of a master equation, i.e., the equation of motion for the reduced density operator describing the quantum state of the system. Master equations are either phenomenologically postulated or derived microscopically from a Hamiltonian model of quantum system plus environment. Contrarily to the case of closed quantum systems, where the equation of motion describing the state dynamics is the Schr\"odinger equation, the general form of the master equation for an open quantum system is not known. Only under certain assumptions, known as the Born-Markov approximation, one can derive a general equation in the so called Lindblad form, able to describe the physical evolution of quantum states~\cite{BreuerPetruccione,Weiss,RivasHuelgaBook,GKSL,Lindblad}. When these assumptions are not satisfied, e.g., for strong system-environment interaction and/or long-living environmental correlations, we enter the intricate (and somewhat fuzzy) reign of non-Markovian dynamics. This consideration already illustrates how, despite its indubitable foundational nature, open quantum systems theory is still far from being completely understood and, in fact, it is peppered with unanswered questions of deep nature.

The increasing ability to coherently control an ever increasing number of individual quantum systems, together with the discovery of quantum coherence in complex biological systems~\cite{HuelgaPlenioQbio}, has  brought to light several scenarios in which the Markovian assumption fails. This has in turn given rise to a proliferation of results on the characterisation of memory effects and non-Markovian dynamics~\cite{RHPReview,BLPReview,InesReview,HallReview}. Interestingly, the cross-fertilisation of ideas from quantum information theory and open quantum systems has led to a new understanding of memory effects in terms of information backflow~\cite{BLPNM,RHPNM,FisherNM,GeometricNM,BognaQChannelNM,NMdegree}, namely a partial return of quantum information previously lost from the open system due to the interaction with the environment.

Experimental results on quantum reservoir engineering, including the possibility to design desired forms of non-Markovian dynamics~\cite{LiuNM2011,Chiuri2011,Mataloni2019,Yu2018,Bernardes2015,Bernardes2016,Cialdi2017,Liu2019,Wittemer2018}, naturally lead to the question of whether or not memory effects are useful for quantum technologies, in the sense of constituting a resource for certain tasks~\cite{Liu2016,BognaQChannelNM,Rossi2018,Laine2014,Chin2012,AddisDD,BognaWork}. This question has not yet been satisfactorily answered. Even more remarkably, a complete understanding of non-Markovianity is still missing, as clearly illustrated in the insightful review of Ref.~\cite{HallReview}.

Given the considerations above, it is not surprising that in recent years a number of experiments have been proposed and realised to verify paradigmatic open quantum system models and test some of their predictions.
Examples are numerous: the milestone experiment on the decoherence of a Schr\"odinger cat state with trapped ions is one of the first examples of engineered environment~\cite{Myatt2000}, followed by the open system quantum simulator of Ref.~\cite{Barreiro2011}.
The latter is also the first experimental realisation of an idea that shifted our perspective about environmental noise. Following a proposal by Vertsrate {\it et al.}~\cite{Verstraete2009,Poyatos}, experimentalists proved that, by engineering certain types of Markovian master equation, one may actually create entangled states as stationary asymptotic states of the dynamics, therefore turning dissipation and decoherence from enemies to allies of quantum technologies~\cite{Barreiro2010,Barreiro2011,Polzik2011}.

In optical platforms, simulators of Markovian open quantum systems have been used to prove the existence of interesting phenomena, such as sudden death of entanglement~\cite{Almeida2007,YuEberly} and sudden transition from quantum to classical decoherence~\cite{Mazzola2010,Xu2010}. In the same platform, experiments have shown how to engineer collisional models~\cite{Mataloni2019}, wherein the microscopic interaction between system and environment is obtained through a sequence of collisions between the open quantum system and one or more ancillae, the latter collectively describing the environment~\cite{Ziman2005}. More recently, experiments in linear optics~\cite{LiuNM2011,Chiuri2011,Bernardes2015, Cialdi2017,Liu2019,Mataloni2019}, cavity QED~\cite{CavityQED}, NMR~\cite{Bernardes2016} and trapped ions~\cite{Wittemer2018} successfully demonstrated the engineering of non-Markovian open quantum systems and monitored the Markovian to non-Markovian crossover. Also complex quantum networks have been proposed as new systems for reservoir engineering of arbitrary spectral densities~\cite{Nokkala2016}, and a bosonic implementation with optical frequency combs has been presented~\cite{Nokkala2018}.

Most of the experiments until now realised for simulating open quantum systems rely on the idea of analogue quantum simulator, that is a quantum system whose dynamics resemble those of another quantum system that we wish to study and understand. In contrast, a digital quantum simulator is a gate-based quantum computer which can be used to simulate any physical system, if suitably programmed~\cite{Feynman,Lloyd}.

Theoretical and experimental research on open systems digital quantum simulators is only now starting to flourish~\cite{NoriPRA2011,TrappedIonsDigital,Wang2013,Lu2017,Xin2017,Shen2017}. While, in principle, general algorithms for digital simulation of open quantum systems have been theoretically investigated~\cite{NoriPRA2011,Wang2013,Shen2017}, their experimental implementation poses several challenges, since the physical quantum gates depend on the experimental platform and the circuit decomposition needs to be optimised in view of gates and measurement errors as well as qubit connectivity. Therefore, the existence of general algorithms for implementing theoretically universal open quantum system simulators does not guarantee the practical implementability in a realistic experiment on a given platform.

In this paper, we demonstrate that a careful circuit decomposition allows us to experimentally implement a vast number of fundamental open quantum systems models for one and two qubits. Not only are we able to generate different classes of open quantum dynamics, namely, unital (e.g., pure dephasing dynamical maps), non-unital (e.g., amplitude damping dynamical maps), phase covariant, and non-phase covariant (e.g., Pauli dynamical maps), but also we can explore the Markovian to non-Markovian crossover, including the recently discovered examples of essential~\cite{NMdegree} and eternal~\cite{Hall2014,Megier} non-Markovianity. We implement a recently proposed non-Markovianity witness~\cite{DetectingDarekChiara} and we use our simulator to prove the non-monotonic behaviour of quantum channel capacity~\cite{ BognaQChannelNM} and extractable work~\cite{BognaWork}, with implications to quantum communication and quantum thermodynamics.

Overall, our results clearly prove that even small quantum processors are versatile and robust testbeds for verifying a number of theoretical open quantum systems results and predictions, therefore paving the way to both new discoveries and a deeper understanding of one of the most fascinating and fundamental fields of quantum physics.

\section*{Results}
We consider an open quantum system represented by a density operator $\rho_S$. The dynamics of the open system are described by a family $\Phi_t$ of completely positive and trace preserving (CPTP) maps, known as the dynamical map: $\rho_S(t) = \Phi_t \rho_S (0)$, with $\rho_S (0)$ the initial state. The equation of motion for the state of the system is the master equation and, if the dynamical map is invertible, can be written in a time-local form
 \begin{equation}\label{}
    \dot{\rho}_S(t) = {\cal L}_t \rho_S (t)\ ,
\end{equation}
where ${\cal L}_t $ is the time-dependent generator of the dynamics:
\begin{equation}\label{Phi-t}
    \Phi_t = {\rm T} \exp\left( \int_0^t {\cal L}_\tau\, d\tau\right)\ ,
\end{equation}
with T the chronological ordering operator, and $\Phi_0= \mathbb{I}$. Under rather general conditions \cite{BreuerPetruccione}, the generator can be written in the form
\begin{eqnarray}\label{GKSL}
    {\cal L}_t \rho_S (t) &=& -i[H_S,\rho_S (t)] \\
    & & +  \sum_k \gamma_k(t) \left( V_k\rho_S(t) V_k^\dagger - \frac 12 \{   V_k^\dagger V_k, \rho_S(t)\} \right). \nonumber
\end{eqnarray}
In the equation above, the first term on the r.h.s.~describes the unitary dynamics, with $H_S$ the system Hamiltonian, and the second term, the dissipative dynamics induced by the interaction with the environment, with $\gamma_k(t)$ and $V_k$ the decay rates and jump operators, respectively. If the decay rates are positive and constant, i.e. $\gamma_k(t)\equiv \gamma \ge 0$, the dynamical map is a semigroup, $\Phi_{t+t'} = \Phi_t \circ \Phi_{t'}$, and we refer to the dynamics as semigroup Markovian. Extending this definition, it is nowadays common to say that the dynamics are Markovian whenever all the decay rates $\gamma_k(t)$ are positive at all times. In this case, the dynamical map satisfies the property of CP-divisibility, namely $\Phi_t = \Phi_{t,s} \circ \Phi_s$, with $\Phi_{t,s}$ a two-parameter family of CPTP maps. Non-Markovian dynamics occur, instead, whenever at least one of the decay rates becomes negative for a certain interval of time. In this case, the intermediate map $\Phi_{t,s} $ is not CP anymore and the dynamics is non-CP-divisible.

In the following subsections, we present experiments run on the IBM Q Experience processors simulating different types of open quantum systems dynamics, both Markovian and non-Markovian.
We begin by considering an example of Markovian semigroup master equation and dynamical evolution.

\subsection*{Markovian reservoir engineering}

\begin{figure*}[t]
  \centering
 \includegraphics[width=\textwidth]{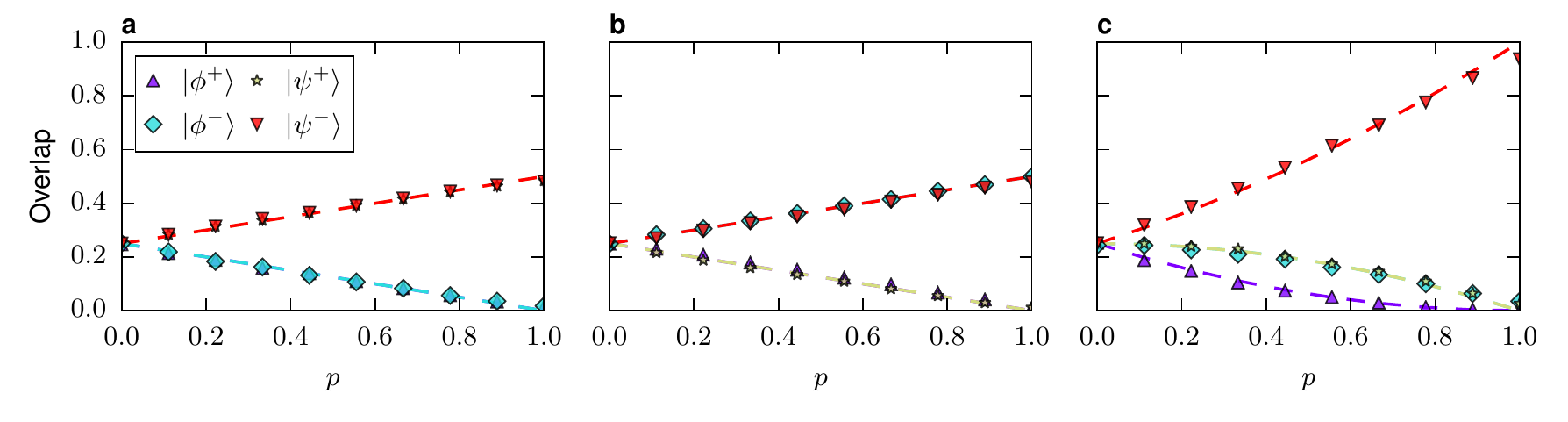}
   \caption{\textbf{Simulation of the Markovian reservoir engineering.} Every plot compares the measured overlap between the state of the system and the four Bell states (dots) with the corresponding analytical prediction for the ZZ pump (\textbf{a}), the XX pump (\textbf{b}) and their consecutive application (\textbf{c}). The pumps are applied to an initial maximally mixed state. The first two pumps clearly reduce the populations of the $+1$ eigenspace of the corresponding stabiliser operator, and their composition results on a probability for the $| \psi^{-} \rangle$ state around 0.94. Throughout the paper, the error bars indicating statistical errors are not shown, as they would be indistinguishable from the markers given the large number of experimental runs (see Methods).}
  \label{fig1}
\end{figure*}

For decades, noise induced by the environment has been considered the archetype enemy of quantum technologies. This is because very often the interaction between a quantum system and its surroundings leads to the fast disappearance of quantum properties, notoriously coherences and entanglement, playing a key role in providing quantum advantage.
This point of view drastically changed as soon as physicists demonstrated that appropriate manipulation of an artificial environment (quantum reservoir engineering) would allow one to steer the open system towards, e.g., a maximally entangled state~\cite{Barreiro2011,Barreiro2010}, hence turning upside down the perspective of the environment as an enemy.

Following the lines of Ref.~\cite{Barreiro2011}, in this subsection we experimentally simulate a semigroup Markovian master equation for a two-qubit open system having as asymptotic stationary state the Bell state $|\psi^- \rangle = \frac{1}{\sqrt{2}}(|01\rangle - |10\rangle)$, where we indicate with $|0\rangle$ and $|1\rangle$ the computational basis of each qubit and we use the notation $| 01\rangle = |0\rangle_1 |1\rangle _2$. This allows us to prepare a maximally entangled state as a result of the dissipative open system dynamics.

 Each of the four Bell states is uniquely determined as an eigenstate with eigenvalues $\pm1$ with respect to $\sigma_z^{(1)} \otimes \sigma_z^{(2)}$ and  $\sigma_x^{(1)}\otimes\sigma_x^{(2)}$, where $\sigma_x^{(i)}$ and $\sigma_z^{(i)}$, with $i=1,2$, are the $x$ and $z$ Pauli operators of qubit 1 and 2. The dissipative dynamics that pumps two qubits from an arbitrary initial state into the Bell state $|\psi^- \rangle $ is realised by the composition of two channels that pump from
the $+1$ into the $-1$ eigenspaces of the stabiliser operators $\sigma_z^{(1)} \otimes \sigma_z^{(2)}$ and  $\sigma_x^{(1)} \otimes \sigma_x^{(2)}$.

Specifically, we consider the two $p$-parametrised families of CPTP maps $\Phi_{zz} \rho_S = E_{1z}  \rho_S  E_{1z}^{\dag} + E_{2z}  \rho_S  E_{2z}^{\dag} $, with
\begin{equation}\label{KrausZZ}
\begin{aligned}
E_{1z} ={}& \sqrt{p} \mathbb{I}^{(1)} \otimes \sigma_x^{(2)} \frac{1}{2}\left( \mathbb{I}+ \sigma_z^{(1)} \otimes \sigma_z^{(2)} \right), \\
E_{2z}  ={}& \frac{1}{2} \left( \mathbb{I}-\sigma_z^{(1)}\otimes\sigma_z^{(2)} \right) \\
&+ \sqrt{1-p} \frac{1}{2} \left( \mathbb{I}+ \sigma_z^{(1)}\otimes\sigma_z^{(2)} \right),
\end{aligned}
\end{equation}
and $\Phi_{xx}\rho_S = E_{1x}  \rho_S  E_{1x}^{\dag} + E_{2x}  \rho_S  E_{2x}^{\dag} $, where $E_{1x}$ and $E_{2x}$ have the same form of  $E_{1z}$ and $E_{2z}$ in Eq.~\eqref{KrausZZ}, provided that we replace $\sigma_x^{(2)}$ with $\sigma_z^{(2)}$ and $\sigma_z^{(1)}\otimes\sigma_z^{(2)}$ with $\sigma_x^{(1)}\otimes\sigma_x^{(2)}$.

By changing the parameter $0 \le p \le 1$ we simulate different types of open quantum system dynamics. For $p\ll1$, the repeated application of, e.g., $\Phi_{zz}$ generates a master equation of Lindblad form with jump operator operator $V=\frac{1}{2}  \mathbb{I}^{(1)}\otimes\sigma_x^{(2)}\left( \mathbb{I}+ \sigma_z^{(1)} \otimes \sigma_z^{(2)} \right)$. For $p=1$, the map $\Phi_{xx} \circ \Phi_{zz}$ generates $|\psi^- \rangle $ for any initial state.

In Fig.~\ref{fig1} we show the action of the dissipative pumping maps $\Phi_{xx}$, $\Phi_{zz}$ and their composition $\Phi_{xx} \circ \Phi_{zz}$ as a function of $p$, for a maximally mixed initial state. We compare the theoretical prediction (dashed lines) with the experimental data. Our results show a very good agreement between theory and experiment for the implementation of both the two families of maps and their composition. In the latter case, the results are more sensitive to errors because of the larger depth of the circuit implementing the composition of maps. Details on the circuit implementations and on their optimisation with respect to the specific qubit connectivity are given in Methods.

\subsection*{Collisional model and essential non-Markovianity}

Our second example of an open quantum system simulator deals with a class of models known as collisional models. These describe the interaction between a quantum system and its environment in terms of consecutive collisions between the system, in our case a qubit, and a sequence of environmental qubits (ancillae) in a given state. The system qubit and the $n$-th environmental qubit interact pairwise during a time period $\tau$.  One assumes, as usual, a factorised initial state of system and environmental qubits. After the system has interacted with a given ancilla, one can trace out the ancilla degrees of freedom as it no longer affects the system's dynamics.

For a sufficiently large number of collisions, when the ancillae are in a thermal state, one can prove that the equation of motion describing the system's dynamics is of Gorini-Kossakowski-Sudarshan-Lindblad (GKSL) form \cite{GKSL}. Interestingly, contrarily to the microscopic derivation of the Markovian master equation, this approach does not rely on Born-Markov approximation, but it automatically leads to a CPTP dynamical semigroup. Also non-Markovian master equations can be introduced in the collisional model picture, for example   by allowing for the ancillae to interact with each other in a well defined manner \cite{ColliMauro,Lorenzo2016,Kretschmer2016}.

We implement experimentally an exemplary model of dynamics displaying the property of essential non-Markovianity \cite{NMdegree}, following the lines of Ref.~\cite{Sergei}. As research in the field of non-Markovian open quantum systems progressed, more refined definitions of memory effects started to appear in the literature. In Ref.~\cite{NMdegree} a hierarchical classification of non-Markovianity was introduced, generalising the notion of CP-divisibility. In particular, the dynamical map is said to be P-divisible if the two-parameter family $\Phi_{t,s}$ is positive. This is, of course, a weaker condition than CP-divisibility, since there exist maps that are P-divisible but not CP-divisible, while all CP-divisible maps are also P-divisible. We say that the dynamics is essentially non-Markovian if it is non-P-divisible, while it is weakly non-Markovian if it is non-CP-divisible but P-divisible.

We consider $n$ ancillae initially prepared in the classically correlated state
$\rho_\textrm{corr} = \frac 12(\ket{0}^{\otimes n}\bra{0}^{\otimes n} + \ket{1}^{\otimes n}\bra{1}^{\otimes n})$. The collision between the system qubit and the $k$-th environmental ancilla is described by the unitary operator $U_k = e^{i g \tau \sigma_z} \otimes |0\rangle_k \langle 0 | +   e^{-i g \tau \sigma_z} \otimes |1\rangle_k\langle 1 |$, where $g$ is the coupling strength. The dynamical map after $n=t/\tau$ collisions is given by
\begin{equation}\label{PhixColl}
\begin{aligned}
\Phi_t \rho_S &= \Tr_E[U_n \cdot \cdot \cdot U_2 U_1 (\rho_S \otimes \rho_\textrm{corr}) U_1^{\dag}  U_2^{\dag} \cdot \cdot \cdot U_n^{\dag} ] \\
&= \cos^2 (ng \tau) \rho_S + \sin^2 (ng\tau) \sigma_z \rho_S \sigma_z.
\end{aligned}
\end{equation}

We compare this dynamics to the case in which the environmental ancillae are prepared in the uncorrelated state $| +\rangle ^{\otimes n} $. In this case, the dynamics is given by
\begin{equation}\label{PhiPColl}
\begin{aligned}
    \Phi^{(+)}_t \rho_S ={} & \frac 12 (1 + \cos^n (2 g \tau) ) \rho_S \\
    & + \frac 12 (1 - \cos^n(2 g \tau)) \sigma_z \rho_S \sigma_z.
\end{aligned}
\end{equation}

\begin{figure}[t]
  \centering
  \includegraphics{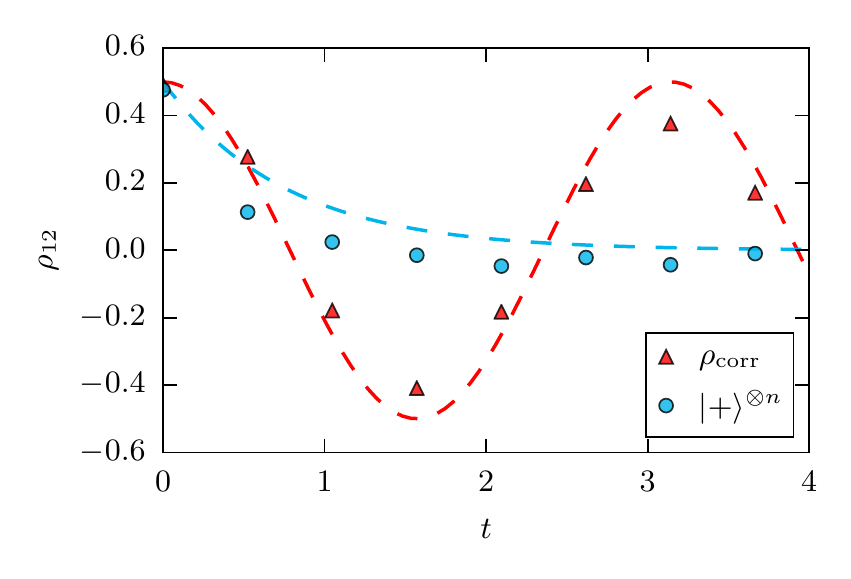}
  \caption{\textbf{Simulation of a collisional model.} The red triangles show the collisions with the ancillae prepared in a classically correlated state (red triangles), while the blue dots show the case of a separable state. In both cases $g=1$, $\tau = \pi/6$ (weak coupling regime). The red and blue dashed curves show, respectively, the theoretical predictions of Eqs.~\eqref{PhixColl} and \eqref{PhiPColl} for time $t=ng\tau$. Up to $7$ collisions were simulated: the depth of the circuits grows with the number of collisions, causing increasing decoherence, but the oscillations due to the essential non-Markovianity are clearly visible in the case of correlated ancillae.}
  \label{coll_model}
\end{figure}

In the weak-coupling case (when $g\tau < \pi / 4$) the map in Eq.~\eqref{PhiPColl} gives rise to Markovian dynamics. In contrast, the map in Eq.~\eqref{PhixColl} alternates intervals, with periodicity $\pi/2$,  in which it is P-divisible ($0<ng\tau<\pi/4$) with intervals in which it is non-P-divisible ($\pi/4<ng\tau<\pi/2$), i.e., it is essentially non-Markovian. In Fig.~\ref{coll_model} we compare, for both maps, the exact dynamics of the qubit coherence with the experimental results, for the initial state $| + \rangle = \frac{1}{\sqrt{2}}(|0\rangle + | 1 \rangle)$. The experimental data are in  good agreement with the theoretical predictions, and the oscillatory behaviour of the coherences in the case of correlated ancillae is a signature of the essential non-Markovianity of the dynamics. We note that the (much smaller) oscillations in the separable case are due to the repeated, imperfect application of the CNOT gates that are required to swap the ancillae that have interacted with the qubit with fresh ones: there is a small probability that the swap fails and the system interacts again with the same ancilla. This can introduce memory effects, as shown, e.g., in Refs. \cite{Lorenzo2016,Kretschmer2016}.

\subsection*{Markovian and non-Markovian dissipative dynamics}

\begin{figure}[t!]
  \centering
  \includegraphics{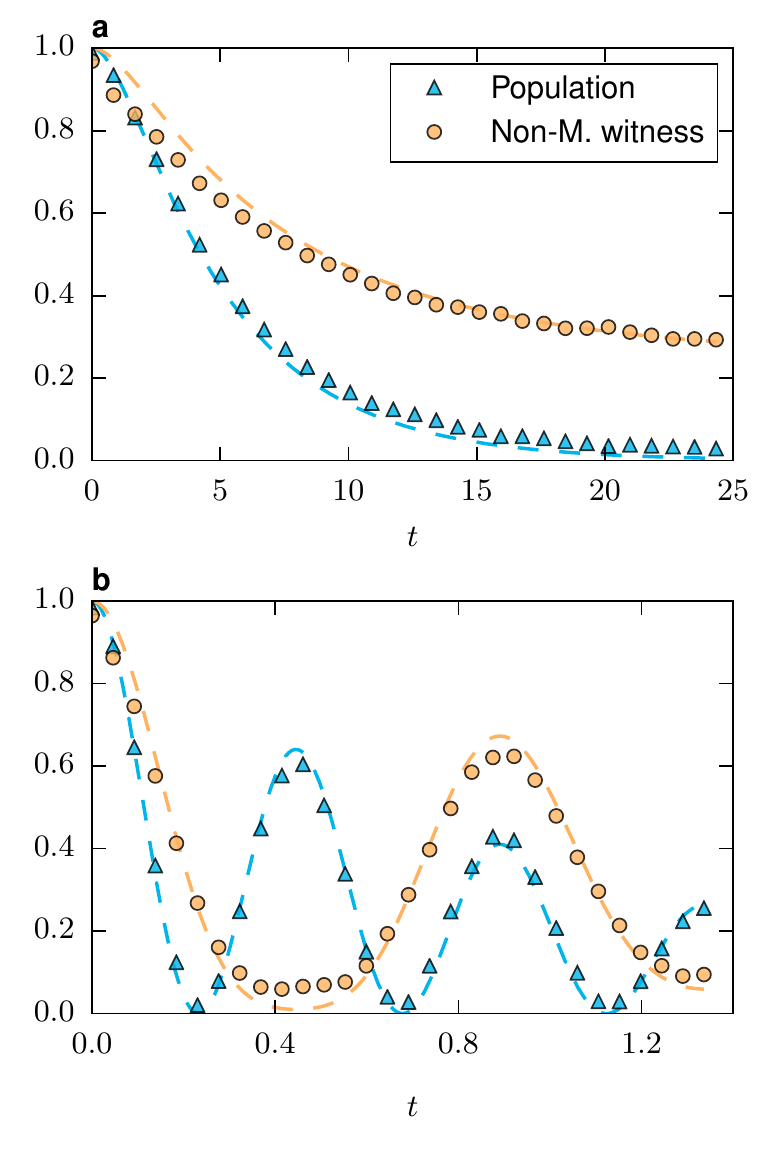}
   \caption{\textbf{Simulation of the amplitude damping channel.} \textbf{a.} Markovian dynamics ($R = 0.2$). \textbf{b.} Non-Markovian dynamics ($R = 100$). Blue triangles show the measured populations of the excited state as a function of time, whereas orange circles stand for the non-Markovianity witness, given by the probability of observing the system and an ancilla, initially prepared in the state $| \phi^{+} \rangle$, in that same state as the channel acts on the system (see Methods for further details). In \textbf{b}, the experimental results clearly show a non-monotonic behaviour, hence confirming the non-Markovian character of the dynamical map. Dashed lines show the theoretical predictions for both quantities.}
  \label{fig3}
\end{figure}

The dynamical maps of Eqs.~\eqref{PhixColl} and \eqref{PhiPColl} are purely dephasing, since they affect only the coherences of the qubit. In this section we simulate an exactly solvable dissipative open quantum system known as the Jaynes-Cummings or generalised amplitude damping model. The microscopic derivation of the master equation can be found in textbooks (see, e.g., Ref.~\cite{BreuerPetruccione}). The master equation is given by
\begin{equation}
\frac{d\rho_{S}}{dt}(t)=\gamma(t)\left[\sigma_{-}\rho_{S}(t)\sigma_{+}-\frac{1}{2}\{\sigma_{+}\sigma_{-},\rho_{S}(t)\}\right],
\label{me}
\end{equation}
where $\sigma_{\pm}$ are the raising and lowering operators and  the time dependent decay rate $\gamma(t)$ has the following analytical form
\begin{equation} \label{ampD}
\gamma(t)=-2\Re\left[\frac{\dot{c}_{1}(t)}{c_{1}(t)}\right],
\end{equation}
with
\begin{eqnarray}
c_{1}(t)&=&c_{1}(0)e^{-\lambda t/2}\left[\cosh\left(\frac{\lambda t}{2}\sqrt{1-2R}\right) \right. \nonumber \\
&& +\left. \frac{1}{\sqrt{1-2R}}\sinh\left(\frac{\lambda t}{2}\sqrt{1-2R}\right)\right].
\label{c1lor}
\end{eqnarray}

The equation above is obtained assuming that the environment is a bosonic zero-temperature reservoir with Lorentzian spectral density
\begin{equation}
J(\omega)=\frac{1}{2\pi}\frac{\gamma_{0}\lambda^{2}}{(\omega_{0}-\omega)^{2}+\lambda^{2}},
\label{lor}
\end{equation}
with $\omega_0$ the qubit frequency, $\gamma_0$ an overall coupling strength, and $\lambda$ the half height width of the Lorentzian profile.
The time-dependent coefficient $c_1(t)$ defined in Eq.~\eqref{c1lor} crucially depends on the ratio $R=\gamma_{0}/\lambda$ between the coupling strength and the width of the spectrum. Also, this coefficient fully determines the qubit dynamics as one can see from the following expression

\begin{equation}
\rho_{S}(t)=
\left(
\begin{array}{cc}
 |c_{1}(t)|^{2} & c_{0}^{*}c_{1}(t) \\
 c_{0}c_{1}^{*}(t) & 1- |c_{1}(t)|^{2}
 \end{array}
\right).
\label{rohs}
\end{equation}

From Eq.~\eqref{c1lor} a straightforward calculation shows that the time-dependent decay rate $\gamma(t)$, defined in Eq.~\eqref{ampD}, takes negative values for certain time intervals whenever $2 R\ge 1$. In this case the dynamical map is not CP-divisible, and therefore non-Markovian. In Fig.~\ref{fig3}\textbf{a} and \textbf{b} we  show, as example of Markovian and non-Markovian dynamics, the evolution of the excited state population of an initially excited state for two values of the parameter $R$, corresponding to Markovian ($R=0.2$) and non-Markovian ($R=100$) dynamics, respectively. The figures show the monotonic decay of the excited state population in the former case, while in the latter case the population oscillates since the qubit exchanges information and energy with the central mode of the Lorentzian peak, resonant with the qubit's Bohr frequency. As done for the other open quantum systems simulators, we compare the experimental data with theoretical predictions, finding a good agreement.

We also implement an experimentally friendly witness of non-Markovianity which was recently introduced in Ref.~\cite{DetectingDarekChiara} and stems from the spectral properties of the dynamical map. The witness
is based on the behaviour of an initial maximally entangled state of qubit and auxiliary ancilla, and requires only the measurement of the expectation values of local observables such as $\sigma_i \otimes \sigma_i$, ($i=x,y,z$). Non-monotonic behaviour of the witness as a function of time signals non-Markovianity. Other witnesses of non-Markovianity (such as the BLP measure introduced in Ref.~\cite{BLPNM}) could be implemented, and may be more efficient at detecting memory effects, but they generally require a larger number of measurements or optimization over initial states.

In Fig.~\ref{fig3}, we plot the dynamics of the entanglement witness for the amplitude damping channel. In Fig~\ref{fig3}\textbf{b}, the witness clearly shows oscillatory behaviour, and therefore properly signals the presence of memory effects. The circuits implementing both the amplitude damping model and the non-Markovianity witness are presented and discussed in Methods.

\subsection*{Depolarizing and Pauli channels}

\begin{figure}[t!]
  \centering
  \includegraphics{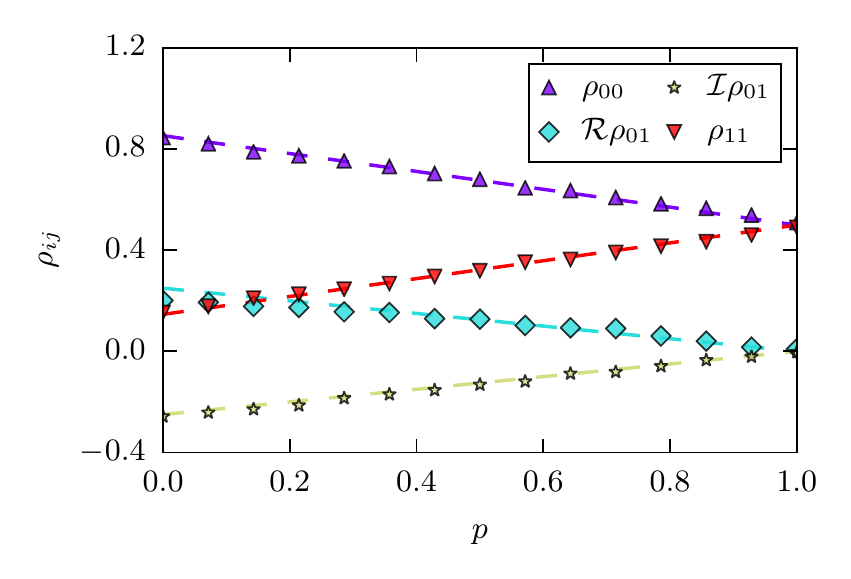}
  \caption{\textbf{Simulation of the depolarizing channel.} Tomography of a single qubit density matrix initially prepared in the state $\ket{\psi_0} = \cos \pi/8 \ket{0} + \sin \pi/8 e^{i \pi/4} \ket{1}$, for various values of the parameter $p$ (cf. Eq.~\eqref{depolarising_channel}).}
  \label{fig:depolarising_channel}
\end{figure}

The depolarizing channel is one of the most common models of qubit decoherence due to its nice symmetry properties. We can describe it by stating that, with probability $p$ the qubit remains intact, while with probability $1-p$ an error occurs. The error can be a bit flip error, described by the action of $\sigma_x$, a phase flip error, described by the action of $\sigma_z$, or both, described by the action of $\sigma_y$. The dynamical map of a Markovian open quantum system subjected to depolarizing noise can be written as
\begin{eqnarray}\label{depolarising_channel}
\Phi_t \rho_S = \left[1-\frac 3 4 p(t)\right] \rho_S + \frac{p(t)}{4} \sum_i \sigma_i \rho_S \sigma_i,
\end{eqnarray}
where $i=x,y,z$ and $p(t)=1 - e^{-\gamma t}$, with $\gamma$ the Markovian decay rate.

In Fig.~\ref{fig:depolarising_channel} we plot the qubit density matrix elements for various values of $p$, comparing the experimental data with the theoretical prediction. For exemplary purposes, we choose a specific initial state possessing non-zero coherences, but we have verified, by repeating the experiment for different initial states, that the agreement observed between experiment and theory is independent from the initial state chosen.
As for the other simulated models, we postpone the discussion on the circuit implementation, the readout, and the error mitigation strategy to the Methods section.

Let us now introduce the most general single-qubit open quantum system model, namely the time-dependent Pauli channel. The master equation in this case takes the form
\begin{equation}
\frac{d\rho_{S}}{dt}(t)=\frac{1}{2}\sum_i\gamma_i(t)\left[\sigma_i\rho_{S}(t)\sigma_i-\rho_{S}(t)\right].
\label{PauliME}
\end{equation}

We note that for $\gamma_i(t) = \gamma$, we recover the Markovian depolarizing channel. Generally, the dynamics described by the master equation above is not phase-covariant \cite{HolevoPC}, except for the case in which $\gamma_x(t)=\gamma_y(t)$. Moreover, since the decay rates may take negative values, conditions for complete positivity must be imposed, and they are given in terms of a set of inequalities involving all the three decay rates, as one can see, e.g., from Ref.~\cite{NMdegree}.

In the Discussion section we present the simulation of a specific form of time-dependent Pauli channel proposed in Ref.~\cite{Hall2014} and used as an example of eternal non-Markovianity, i.e.~an open quantum system dynamics for which the dynamical map is non-CP-divisible for all times $t$. More precisely, we use this experimental simulation to demonstrate a phenomenon predicted in Ref.~\cite{BognaWork}, namely the presence of oscillations in the extractable work. This shows an application of open quantum system simulation on the IBM Q Experience processors to fields other than quantum information theory, specifically quantum thermodynamics for the example here considered.

\section*{Discussion}

The results presented in the previous section highlight how few-qubits NISQ devices publicly available on the cloud already provide sufficient robustness and reliability to implement experimentally a number of open quantum systems dynamics, both Markovian and non-Markovian, both unital (dephasing collisional model) and non-unital (amplitude damping, depolarizing, and Pauli channels).

We have simulated all the paradigmatic open quantum systems models typically used to demonstrate physical phenomena induced by the presence of the environment, its consequences for quantum technological applications, but also possible benefits in the spirit of reservoir engineering. As an example of a fundamental study on open quantum systems, we have measured the dynamics of a non-Markovianity witness and showed that it correctly signals the presence of memory effects for time-local amplitude damping dynamics.

In this section we build on these results to substantiate our claim of the IBM Q Experience being a versatile testbed for the experimental verification of physical effects due to the open character of the dynamics. Specifically, we focus on two applications: non-Markovian quantum channel capacity, of potential use in quantum communication, and extractable work dynamics, relevant in quantum thermodynamics.

\begin{figure}[t!]
  \centering
  \includegraphics{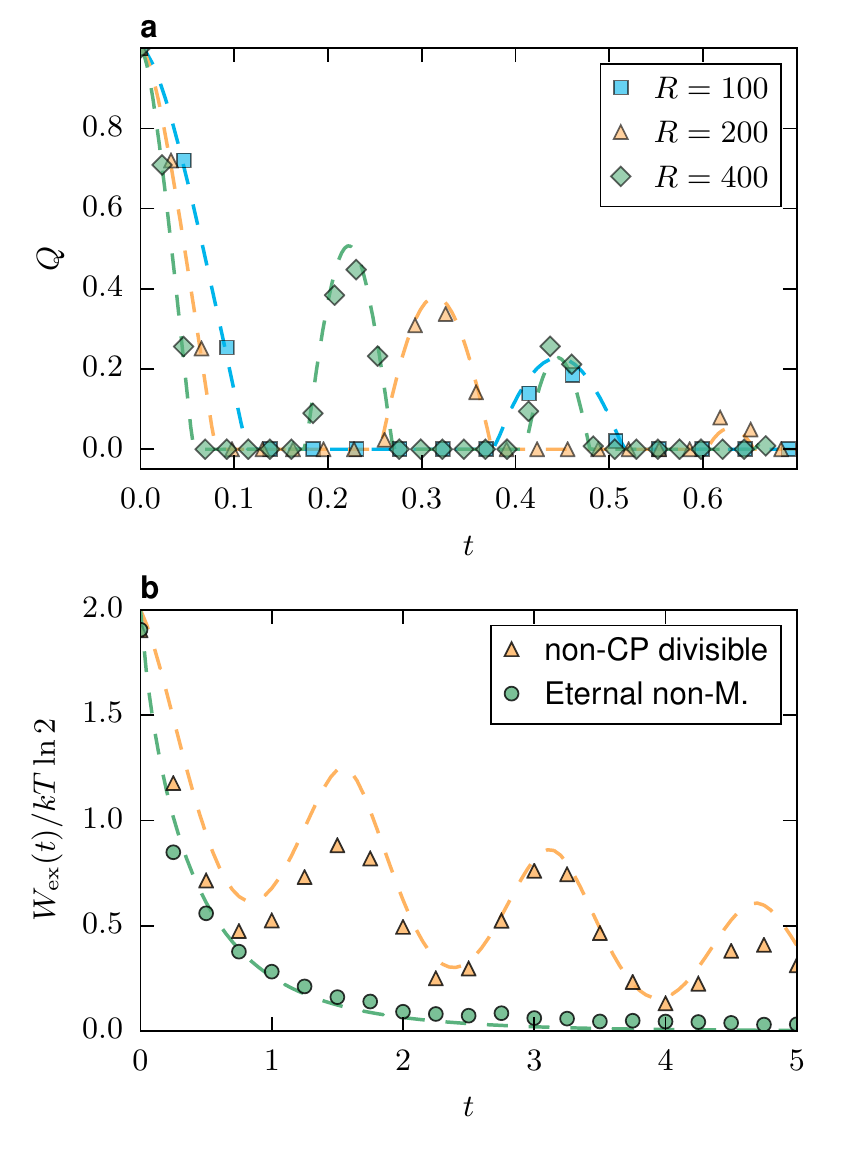}
  \caption{\textbf{Applications in quantum communication and thermodynamics.} \textbf{a.} Quantum channel capacity as a function of time for three different values of $R$ calculated from Eq.~\eqref{eq:channel_cap}, where $|c_1(t)|^2$ has been approximated by the ratio between the population of the excited state at time $t$ and at $t=0$ in order to account for the deviations in the initial state preparation. \textbf{b.} Rescaled extractable work $W_{\mathrm{ex}}(t)/k T \ln 2$ as a function of time for the eternally non-Markovian Pauli channel, with $\gamma_x(t)=\gamma_y(t)=\lambda/2$ and $\gamma_z(t)=-\omega \tanh \omega t  / 2$, with $\lambda = 1$ and $\omega = \frac 12$ (green circles) and for the non-CP-divisible Pauli channel $\gamma_x(t)=\gamma_y(t)=\lambda/2$ and $\gamma_z(t)=\omega \tan \omega t / 2$, with $\lambda = 0.1$ and $\omega = 2$. The extractable work has been evaluated using Eq.~\eqref{workext} from a full two-qubit tomography of system $S$ and memory ancilla $M$.}
  \label{fig:applications}
\end{figure}

Let us first consider a typical setting in quantum information processing and communication: Alice and Bob are at the opposite ends of a quantum channel, the former sending information (classical or quantum) and the latter receiving it. The maximum amount of information that can be reliably transmitted along a noisy quantum channel is known as the channel capacity. Specifically, we consider the quantum capacity $Q$ defined as the limit to the rate at which quantum information can be reliably sent down a quantum channel.

In the following, we focus on the time-local amplitude damping model introduced in the previous section. In this case, the quantum channel capacity can be calculated exactly and takes the form \cite{BognaQChannelNM}
\begin{equation}\label{eq:channel_cap}
\begin{aligned}
Q(\Phi_t) = \max_{p\in[0,1]} & \left[ H_2 (|c_1(t)|^2 p) - \right. \\
& \left. H_2 ([1-|c_1(t)|^2] p) \right],
\end{aligned}
\end{equation}
for $|c_1(t)|^2>1/2$ and $Q(\Phi_t)=0$ otherwise. In the equation above  $H_2$ is the binary Shannon entropy and $c_1(t)$ is given by Eq.~\eqref{c1lor}.

One may be led into believing that the quantum channel capacity decreases, and in some case disappears, for increasing lengths of the noisy channel, due to the cumulative destructive effect of decoherence. This, however, only holds for Markovian noise, as theoretically demonstrated in Ref.~\cite{BognaQChannelNM}. Indeed, due to memory effects, the quantum channel capacity may take again non-zero values after disappearing for a certain finite length of the channel. In Fig.~\ref{fig:applications}\textbf{a} we experimentally demonstrate this effect, comparing the measured behaviour of $Q(\Phi_t)$ with the theoretical prediction for an amplitude damping model with $R=100$, $R=200$, and $R=400$.

We now consider an example of interest in the field of quantum thermodynamics, specifically related to the presence of memory effects in the open system dynamics. In Ref.~\cite{BognaWork} the connection between information back-flow and thermodynamic quantities was established by means of a non-Markovianity indicator based on the quantum mutual information. More specifically, it was proven that there exists a link between memory effects, as indicated by oscillations in the quantum mutual information and the non-monotonic behaviour of the extractable work $W_{\mathrm{ex}}$.

To understand this connection, let us recall that in order to link non-Markovianity with the evolution of thermodynamical quantities one needs to describe not only the system $S$, whose information content we are interested in, but also an observer or memory $M$. Interestingly, quantum mechanical correlations between system and observer may lead to the exciting possibility of extracting work while erasing information on the system \cite{Oscar}.

Following Ref.~\cite{BognaWork},  we consider the case of a qubit $S$ subjected to a Pauli dynamical map, see master equation \eqref{PauliME}, and an isolated qubit acting as memory, $M$. If the system and memory are prepared in a maximally entangled state, work can be extracted during the information erasure by using the initial entanglement, as predicted in Ref.~\cite{Oscar}. In this framework, the extractable work takes the form  \cite{BognaWork}:
\begin{eqnarray}\label{workext}
W_{\mathrm{ex}} (t) = \left[n - H(\Phi_t \rho_S) +  I(\Phi_t \rho_S: \rho_M)\right] kT \ln 2,
\end{eqnarray}
with $n$ the number of qubits in $S$, $H(\Phi_t \rho_S)$ the von Neumann entropy of the evolved state of the system, $I(\Phi_t \rho_S: \rho_M)$ the quantum mutual information quantifying the amount of total correlations between system $S$ and memory $M$, $k$ the Boltzmann constant and $T$ the temperature of the reservoir used for the erasure. For the open system here considered  revivals of extractable work are due to the interplay between memory effects, witnessed by the quantum mutual information, and entropy dynamics, as dictated by Eq.~\eqref{workext}.

We now further specify our analysis to the eternal non-Markovianity model, for which $\gamma_x (t) = \gamma_y (t) = \lambda/2$ and $\gamma_z(t)= - \omega \tanh (\omega t)/2$. We compare the dynamics with the case in which $\gamma_x (t) = \gamma_y (t) = \lambda/2$ and $\gamma_z(t)=  \omega \tan (\omega t)/2$. The first dynamics is eternally non-Markovian, since $\gamma_z(t) < 0$ at all times, while the second one is non-CP-divisible but not eternally non-Markovian, since  $\gamma_z(t)$ takes both positive and negative values. This comparison allows us to distinguish between quantum and classical memory effects~\cite{BognaWork}.

In Fig.~\ref{fig:applications}\textbf{b} we show the experimental results and the theoretical prediction for the dynamics of the extractable work for the two cases here considered. Despite the presence of experimental imperfections, clear oscillations of the extractable work are visible in one case, but completely absent in the other one, therefore showing for the first time not only the impact of memory effects on a thermodynamic quantity such as $W_{\mathrm{ex}}$ but also the subtle origin of these oscillations.

To conclude, it is worth mentioning that experiments on open quantum systems performed in dedicated experimental laboratories in different platforms such as, e.g., trapped ions, linear optical systems, superconducting qubits, NMR, NV centers in diamonds, cavity QED, etc., generally reach higher precision and fidelity than those reported in our paper. However, this often occurs at the expense of increased specificity in terms of the models that can be simulated. Moreover,  these experiments are accessible only to the researchers working in the given laboratory.

Open source small quantum processors already are, however, a reality and they have opened the door of experimental physics also to quantum researchers who do not have either specific experimental training or sufficient economic resources to set up a laboratory. We therefore envisage a rapid change in the way in which research is conducted in a field which has been until now dominantly theoretical, namely the study of open quantum system dynamics. Perhaps, in a not too far future, this will blur the line separating theoretical and experimental research, and an increasing number of experiments will be remotely programmable by using a simple interface and language,  in the true spirit of a quantum simulator.



\section*{Methods}

The results presented in this paper have been obtained on the IBM Q Experience processors freely available online: two 5-qubit machines, \texttt{ibmqx2} and \texttt{ibmq\_vigo}, and a 14-qubit machine, \texttt{ibmq\char`_16\char`_melbourne}.
The circuits have been written using IBM's Qiskit~\cite{qiskit}, a Python-based programming language for the IBM Q Experience.
Each circuit has been run for 8192 shots (the maximum allowed by the devices) to gather statistics from the measurements.
When needed, one- and two-qubit full state tomography has been obtained by using the tools provided in Qiskit, performing a maximum-likelihood reconstruction of the density matrix \cite{Smolin2012}.

In the following, we give detailed explanations of the various circuits implemented to produce the results presented in the paper. We first make a few general considerations on the sources of error and on how to devise circuits with high fidelity on the IBM Q Experience devices.

Given the large number of experimental runs allowed by the devices, the errors due to statistical fluctuations, of order $\mathcal{O}(1/\sqrt{N}) \approx 0.011$, are too small for errorbars to be distinguishable from markers in Fig.~\ref{fig1}-\ref{fig:applications}, and thus they are not shown. The discrepancy between the experimental points and the theoretical prediction, on the other hand, comes from systematic errors induced by various factors in the hardware implementation. First, the qubits undergo relaxation and decoherence due to external noise. This error becomes more relevant when increasing the depth of the circuit. Second, the gates have errors due to cross-talk and unwanted interactions between qubits when addressing them with pulses. In particular, the error rate of CNOT gates is about 10 times larger than single-qubit unitaries. Finally, there is a readout error affecting the quantum measurement, although this can be mitigated to some extent, as we discuss later on. The error rates and noise parameters on IBM Q devices are characterized on a daily basis using randomized benchmarking techniques. Since there are various, interdependent sources of noise, modelling and predicting the deviations in the experimental data is a non-trivial matter, beyond the scope of this paper. In this work, all
quantum channels have been decomposed into circuits bearing in mind some
general guidelines targeted at minimising the aforementioned inaccuracies.

Since the IBM Q Experience devices are universal quantum computers, they enable
the implementation of any unitary transformation of their constituent qubits;
once a quantum circuit is provided for its execution, it is compiled into an
equivalent circuit involving only the machine's basis gates, that is, those
realisable experimentally. In the case of the IBM Q Experience devices, these include any single-qubit rotation, whereas the only multi-qubit gates are CNOTs. However, if the circuit requires a multi-qubit gate
among qubits that are not physically connected, the corresponding gate will be replaced  with a
longer circuit in which the states are swapped to neighbouring qubits in the
compiled circuit. Since every swap gate includes a minimum of three CNOT gates,
and these introduce considerable noise to the execution, it is crucial to assign
the relevant qubits involved in the simulation (e.g.~system, environment and
ancillae) to the machine's qubits so that the number of CNOT gates between
disconnected qubits is minimised. Furthermore, the devices are calibrated daily
and the errors of the basis gates are reported. This information can also be
taken into account in the qubit assignment, as using the CNOT gates with smaller
errors is preferable.

In addition to the gate errors, the qubits' readout errors characterising the discrepancies between the qubit state and
the measurement outcome probabilities are also provided. In the IBM Q Experience devices, there are considerable
differences in the readout errors of the different qubits, so this information
should also be taken into account in the qubit-assignment process: if possible,
it is preferable to assign the system (and any auxiliary ancillae whose measurement
is required) to low readout-error qubits, while qubits with large readout errors
can still be used to simulate the environment or other ancillae that need not be measured. In any case, Qiskit provides
post-processing error-mitigation tools that generally improve the experimental
results under the package \texttt{ignis}. To do so, we first prepare all possible basis states
$\ket{0\cdots 0}, \ket{0\cdots 1}, \ldots, \ket{1\cdots 1}$ and
measure their corresponding outcome probabilities (notice that it is
only necessary to include the qubits whose measurement outcomes need to be
corrected). Once these are known, they can be used to correct any other
experimental result by finding, via likelihood maximisation, the experimental
outcome that is most congruent with the observed measurement deviations. All the data shown in the paper have been mitigated as described above, with the only exception of the collisional model with the ancillae prepared in the separable state; the reason why we have not mitigated those results is that, due to the qubit swapping, the measurements are performed on different physical qubits.

\subsection*{Reservoir engineering}

\begin{figure*}[t]
  \centering
  \includegraphics{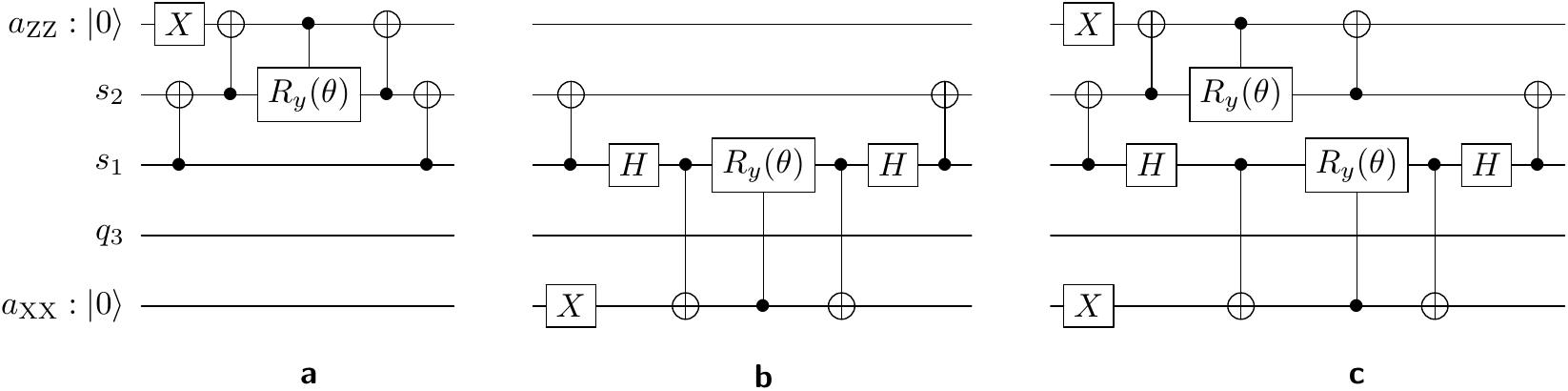}
  \caption{\textbf{Circuits implementing the reservoir engineering protocol}. \textbf{a.} ZZ pump, \textbf{b.} XX pump and \textbf{c.} ZZ
  and XX pump. The circuits were run on \texttt{ibmqx2}.
  Qubits $s_1$ and $s_2$ (corresponding to $q_2$ and $q_1$ on the device) are
  the system qubits, while qubits $a_\textrm{XX}$ and $a_\textrm{ZZ}$ ($q_4$ and $q_0$) are
  the environment ancillae for the two maps. State preparation and measurement
  are not shown in the circuits above.}
  \label{fig:reservoir_engineering}
  \end{figure*}

    In Ref.~\cite{Barreiro2011}, the authors provide the circuits for the implementation of the Bell-state pumping. However, these are composed of gates that are natural to the trapped-ions platform used in that work, so their direct implementation on the IBM Q Experience devices would result in far too long circuits. Therefore, we propose a different set of circuits that follow the same basic working principles, but have been designed specifically keeping in mind the characteristics of the IBM Q Experience platform.

    The pumping circuits proposed in Ref.~\cite{Barreiro2011} are composed of four parts. First, the relevant information regarding the state of the system (that is, whether the system is in the $+1$ or the $-1$ eigenspaces of the stabiliser operators) is mapped into an ancilla. Second, the state of the system is modified depending on the state of the ancilla. Third, the mapping circuit is reversed. At this stage, the system has been pumped, but if the ancilla is to be used again for a new pumping cycle, it needs to be reset, which is the fourth step. We follow these same lines, designing circuits that perform these same steps while minimising the number of gates involved. Before we explain the resulting circuits, let us mention that, since the IBM Q Experience devices are not equipped with the reset operation, we must use a different ancilla for every pump.

The way we map the eigenspace information into an ancilla is by first applying a CNOT gate between the system qubits. Suppose that qubits $s_1$ and $s_2$ are initially in some Bell state, for instance, $| \phi^{\pm} \rangle = (| 00 \rangle \pm | 11 \rangle)/\sqrt{2}$. A CNOT gate controlled by $s_1$ transforms the state into $|\pm\rangle|0\rangle$. Instead, $| \psi^{\pm} \rangle$ would be transformed into $|\pm\rangle|1\rangle$. Hence, we see that the information regarding the $\sigma_x^{(1)}\otimes\sigma_x^{(2)}$ eigenspace (namely, the sign) is contained in the state of $s_1$ after the transformation, whereas the one corresponding to the $\sigma_z^{(1)}\otimes\sigma_z^{(2)}$ is in qubit $s_2$. Now, let us consider the circuit implementing the $\sigma_z^{(1)}\otimes\sigma_z^{(2)}$ pump in Fig.~\ref{fig:reservoir_engineering}\textbf{a}. To map the eigenspace information into the environment ancilla $a_\textrm{ZZ}$, we apply a CNOT controlled by the relevant qubit, $s_2$. After these two gates (and considering that the initial state of the ancilla is $|1\rangle$), $a_\textrm{ZZ}$ will be in state $|1\rangle$ if the initial state of the system is $| \phi^{\pm} \rangle$ and $|0\rangle$ if it is $| \psi^{\pm} \rangle$. Therefore, the conditional rotation gate only acts in the former case, while it does not modify the state in the latter. The angle of the controlled rotation, in turn, controls the efficiency of the pump $p$ via the relation $\theta = 2 \arcsin{\sqrt{p}}$. Finally, the last two CNOT gates simply revert the mapping part of the circuit. The working principle of the $\sigma_x^{(1)}\otimes\sigma_x^{(2)}$ pump (Fig.~\ref{fig:reservoir_engineering}\textbf{b}) is essentially the same. However, we need to add an extra Hadamard gate to transform the state of $s_1$ before mapping the information to the ancilla $a_\textrm{XX}$. As for the composite pump, we can simply concatenate the two circuits. Notice that in the direct concatenation there would be two consecutive CNOTs between the system qubits, which can be removed.

    Regarding the measurement process, the IBM Q Experience platform only enables measurement in the computational basis. Hence, to assess the probabilities for each of the Bell states, we need to change basis by applying again a combination of CNOT and Hadamard to the system qubits, so $| \phi^{+} \rangle$ is mapped into $|00\rangle$, etc. Again, notice that this would result in repeated consecutive gates, the effect of which amounts to identity, so they can be removed from the circuits. Finally, in Fig.~\ref{fig1} we show the results starting from the maximally mixed state. To do so, we simulated the circuits preparing the system in four different initial conditions, $| 00 \rangle$, $| 01 \rangle$, $| 10 \rangle$ and $| 11 \rangle$.

\subsection*{Collisional model}

\begin{figure*}[t]
  \centering
  \includegraphics{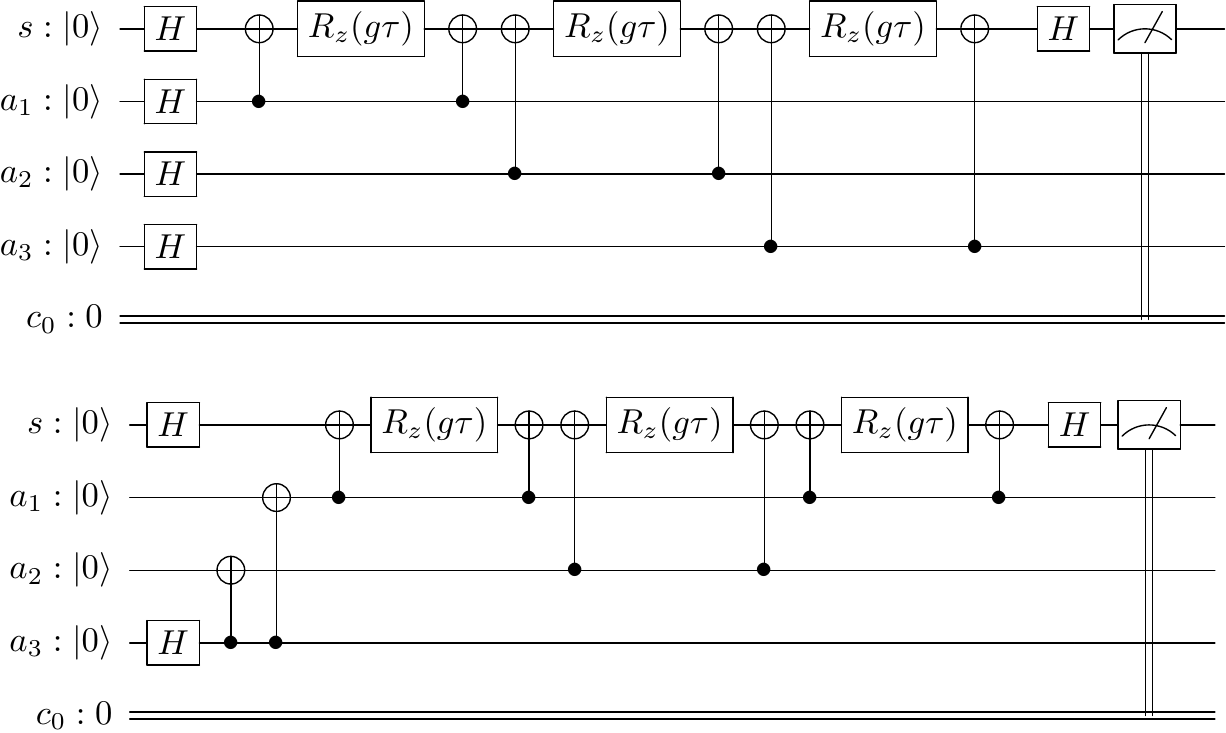}
  \caption{\textbf{Circuits implementing the collisional model}. The top circuit shows the case of the collision with 3 ancillae in the separable $\ket{+++}$ state, prepared by means of Hadamard gates. Each collision consists of a rotation around $Z$ between two CNOTs. A final measurement in the $X$ basis is performed in order to measure the coherence of the system qubit. The bottom circuit shows how, in the case of ancillae prepared in the correlated state, we can effectively simulate the map with three ancillae prepared in the GHZ state $(\ket{000} + \ket{111})/\sqrt{2}$, by colliding alternately with two of them. Notice that, in principle, we could have always a collision with the same ancilla. This, however, would cause the compiler to remove consecutive CNOTs and join the rotations into a single gate, making the circuit trivial. The top circuit was run on \texttt{ibmq\char`_16\char`_melbourne}, while the bottom circit was run on \texttt{ibmqx2}.}
  \label{fig:circuit_collisional_model}
\end{figure*}

The circuit used to implement the collisional model described in section ``Collisional model and essential non-Markovianity''
is depicted in Fig.~\ref{fig:circuit_collisional_model}. Initially, the
system and ancillae are prepared in the desired initial state, then each
collision is applied, and finally the qubit is measured in the $\sigma_x$
basis. The collision unitary $U = e^{i g \tau \sigma_z} \otimes \ket{0}\bra{0} +
e^{-i g \tau \sigma_z} \otimes \ket{1}\bra{1}$ can be implemented by means
of a rotation $R_z(g \tau)$ around the $Z$ axis between two CNOT gates.

When the ancillae are prepared in the state $\rho_\textrm{corr} = (\ket{0}^{\otimes n}\bra{0}^{\otimes n} + \ket{1}^{\otimes n}\bra{1}^{\otimes n})/2$, we can simplify the circuit by noticing that each ancilla is in the maximally mixed state, and the action of $U$ does not affect its state. We thus only need three ancillae prepared in the GHZ state $\ket{\psi_\textrm{GHZ}} = (\ket{000}+\ket{111})/\sqrt{2}$: by tracing out the third one we are left with the two-qubit state $\rho^{(2)}_E = (\ket{00}\bra{00} + \ket{11}\bra{11})/2$, and then we can keep colliding with either of the two ancillae.

The circuit with the ancillae in the separable state was implemented and run on the device \texttt{ibmq\char`_16\char`_melbourne} ($14$ qubits), in order to have a larger number of ancillae for the collisions. The system qubit was chosen to have the highest connectivity (3 qubits) and smallest readout error. The connectivity layout of the computer does not allow for direct collisions with more than $3$ ancillae, and swaps between qubits are required, increasing the errors in the simulation and possibly introducing memory effects (as discussed in the Results section).

\subsection*{Amplitude damping channel}

\begin{figure}[t]
  \centering
 \includegraphics[]{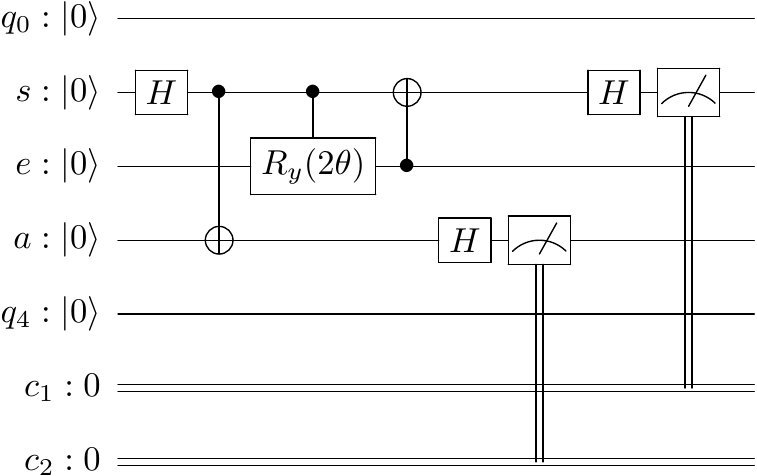}

  \caption{\textbf{Circuit for the amplitude damping channel with non-Markovianity witness.} The circuit was run on \texttt{ibmq\_vigo}. Qubits $q_1$, $q_2$ and $q_3$ were used as system, environment, and witness ancilla, respectively. Setting  $\theta = \arccos{c_1(t)}$, the channel acting on the system qubit simulates the amplitude damping dynamics. The two Hadamard gates rotate the state of the qubits in order to measure the observable $\sigma_{x}\otimes\sigma_{x}$. To measure $\sigma_{y}\otimes\sigma_{y}$, the combination $S^\dagger H$ can be used instead, whereas no gates are required for $\sigma_{z}\otimes\sigma_{z}$.}
  \label{fig:amplitude_damping_circuit}
\end{figure}{}

The circuit in Fig.~\ref{fig:amplitude_damping_circuit} implements the amplitude damping channel with the non-Markovianity witness. For an arbitrary pure state of the system $|\psi\rangle_s = \alpha |0\rangle_s + \beta |1\rangle_s$, and setting the state of the environment to vacuum $|0\rangle_e$, the two gates between the system and environment qubits transform the joint state into $\alpha |0\rangle_s|0\rangle_e + \beta \cos \theta |1\rangle_s|0\rangle_e + \beta \sin \theta |0\rangle_s|1\rangle_e$. Therefore, identifying the states $|0\rangle_s$ and $|1\rangle_s$ with the ground and excited states respectively, and by choosing $\theta = \arccos{c_1(t)}$, the reduced state of the system becomes Eq.~\eqref{rohs}.

The non-Markovianity witness for a channel $\Phi_t$ proposed in Ref.~\cite{DetectingDarekChiara} is based on the dynamical behaviour of the entanglement between the system and an ancilla initially prepared in a maximally entangled state. Namely, this quantity is defined as $f_{\Phi} = \langle \phi^{+} | \mathbb{I} \otimes \Phi_t [| \phi^{+} \rangle \langle \phi^{+} | ] | \phi^{+} \rangle$. Hence, implementing this quantity requires preparing the state $| \phi^{+} \rangle$ between the system and an ancilla, which can be achieved using a Hadamard and a CNOT gates, applying the dynamical map to the system and measuring the probability of finding the joint system and ancilla state in $| \phi^{+} \rangle$. As suggested in Ref.~\cite{DetectingDarekChiara}, we need not use an extra CNOT gate to project on the Bell basis; instead, we can take advantage of the fact that $| \phi^{+} \rangle \langle \phi^{+} | = (\mathbb{I}\otimes \mathbb{I} + \sigma_{x}\otimes \sigma_{x} - \sigma_{y}\otimes \sigma_{y} + \sigma_{z}\otimes \sigma_{z})/4$ and measure the corresponding local observables. Figure \ref{fig:amplitude_damping_circuit} shows the circuit corresponding to the measurement of $\sigma_{x}\otimes \sigma_{x}$.

\subsection*{Depolarizing channel}

\begin{figure}[h]
  \centering
 \includegraphics{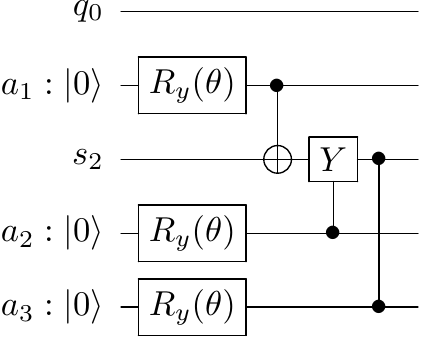}
  \caption{\textbf{Circuit implementation of the depolarizing channel}. The circuit was run on \texttt{ibmqx2}.
      Qubit $q_2$ is used as the system qubit, as it is the
       only one that can be targeted with three CNOTs. The ancillae $a_1,a_2$ and $a_3$ are initially in the state $\ket{0}$ and rotated around the $y$ direction by an angle $\theta(p) = \frac 12 \arccos(1-2p)$. They then act as controls for controlled-$X$, -$Y$ and -$Z$ gates, respectively.}
  \label{fig:depolarising_channel_circuit}
\end{figure}

The depolarizing channel defined in Eq.~\eqref{depolarising_channel} can be implemented, for any value of $p\equiv p(t) \in [0, 1]$, with the circuit shown in Fig.~\ref{fig:depolarising_channel_circuit}.
Three ancillary qubits are prepared in a state $\ket{\psi_\theta} = \cos \theta/2 \ket{0} + \sin \theta/2 \ket{1}$,
and are used as controls for, respectively, a controlled-$X$ (CNOT), a controlled-$Y$ and a controlled-$Z$ rotation.
This way, each gate will be applied with a probability $\sin^2 \theta/2$.

The rotation angle $\theta$ must be chosen so that each of the gates is
applied with probability $p$. Notice that applying $X$ and then $Y$,
but not applying $Z$ is equivalent (up to global phases) to just applying $Z$, and so on. The resulting equation that binds $\theta$ to $p$ is thus
\begin{equation}
    \sin^2 \frac \theta 2 \cos^4\frac\theta2 + \sin^4 \frac\theta 2 \cos^2 \frac \theta 2 = \frac p 4,
\end{equation}
with solution $\theta(p) = \frac 12 \arccos(1 - 2 p)$.

\subsection*{Pauli channel}

\begin{figure}[h]
  \centering
  \includegraphics{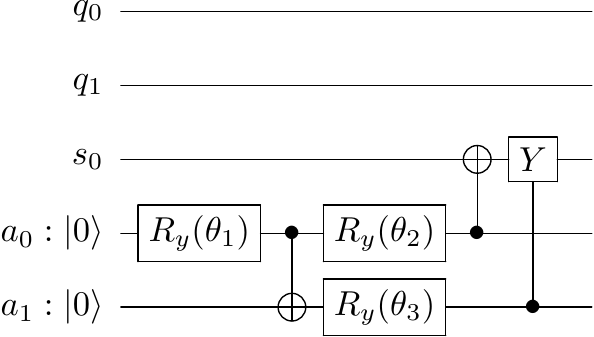}
  \caption{\textbf{Circuit implementation of a general Pauli channel on \texttt{ibmqx2}.}
      Qubit $q_2$ is used as the system qubit because it can be targeted
      with two CNOTs and has the smallest readout error. The ancillae are initially in the state $\ket{0}$ and rotated around $y$ with
      the angles
      $\theta_1, \theta_2$ and $\theta_3$
      found by solving the system in Eq.~\eqref{eq:pauli_equations}, and they act as controls for a controlled $X$ and a controlled $Y$.}
       \label{fig:circuit_pauli}
\end{figure}

At a specific time instant $t$, the Pauli channel described by the master equation \eqref{PauliME} can be written as
    \begin{equation}
        \mathcal{E} (\rho) = \sum_{i=0}^3 p_i \sigma_i \rho \sigma_i,
    \end{equation}
    with $0 \leq p_i \leq 1$ and $\sum_i p_i = 1$. The depolarizing channel
    is a special case of the Pauli channel where $p_1 = p_2 = p_3 = p/4$.

    One could think to implement the channel by generalising the circuit of Fig.~\ref{fig:depolarising_channel_circuit}, specifically, by allowing for the three ancillae to be prepared in different states. One can see, however, that this cannot be done in the most general case. For example, the eternally non-Markovian channel presented in the Discussion section is not implementable in this way.

    It is possible to implement the general Pauli channel with just two qubits, if we allow them to be in an entangled state. The first qubit acts as the control for a controlled-$X$ (CNOT) gate, and the second one for a controlled-$Y$.
    Notice that, as remarked in the previous section, applying both a controlled-X and a controlled-Y is effectively equivalent to applying a controlled-Z.

    The state $\ket{\psi}$ of the ancillae needed for the Pauli channel can be implemented by the circuit in Fig.~\ref{fig:circuit_pauli}, parametrised by the three angles $\theta_1,\theta_2,\theta_3$:
    \begin{equation}
        \begin{aligned}
        \ket{\psi} =& (c_1 c_2 c_3 + s_1 s_2 s_3) \ket{00} + \\
                    & (c_1 c_2 s_3 - s_1 s_2 c_3) \ket{01} + \\
                    & (c_1 s_2 c_3 - s_1 c_2 s_3) \ket{10} + \\
                    & (s_1 c_2 c_3  + c_1 s_2 s_3) \ket{11}
        \end{aligned}
    \end{equation}
    where $c_i \equiv \cos \theta_i$ and $s_i \equiv \sin \theta_i$.

    The angles $\theta_i$ can be found by solving the following system of equations:
    \begin{equation}\label{eq:pauli_equations}
    \begin{cases}
        p_0  = |\braket{00|\psi}|^2 = (c_1 c_2 c_3 + s_1 s_2 s_3)^2 & \\
        p_1  = |\braket{01|\psi}|^2 = (c_1 c_2 s_3 - s_1 s_2 c_3)^2 & \\
        p_2  = |\braket{10|\psi}|^2 = (c_1 s_2 c_3 - s_1 c_2 s_3)^2 & \\
        p_3  = |\braket{11|\psi}|^2 = (s_1 c_2 c_3 + c_1 s_2 s_3)^2 &
    \end{cases}
    \end{equation}

The above system of equations allows for multiple analytical solutions whose expressions are too cumbersome to be reported here. The choice of the solution to use in each case depends on a number of factors, such as the gate fidelity for the specific values of the parameters.

Notice that the circuit for the Pauli channel can be used also for the depolarizing channel, instead of the three-qubit circuit of Fig.~\ref{fig:depolarising_channel_circuit}, but it proves to be less accurate on the IBM Q Experience devices, presumably because of the required entanglement between the two qubits.

\section*{Data availability}
All data supporting the findings of this study are available from the authors upon request. The code to reproduce all the results is available online at \url{https://github.com/matteoacrossi/ibmq_open_quantum_systems}.

\section*{Acknowledgements}
 We thank Dariusz Chru\'{s}ci\'{n}ski and Thomas Bullock for interesting discussions. We acknowledge use of the IBM Q Experience for this work. The views expressed are those of the authors and do not reflect the official policy or position of IBM or the IBM Q team.
 We acknowledge financial support from the Academy of Finland via the Centre of Excellence program (Project no.~312058 as well as Project no.~287750). G.~G.-P. acknowledges support from the emmy.network foundation under the aegis of the Fondation de Luxembourg.

 \section*{Competing interests}
 The authors declare no competing interests.

\section*{Author contribution}
  G.~G.-P., M.~A.~C.~R.~and S.~M.~contributed to the design and implementation of the research, to the
analysis of the results, and to the writing of the manuscript. G.~G.-P.~and M.~A.~C.~R.~are co-first authors.

\end{document}